\documentclass[a4paper,10pt,twoside]{cpc-hepnp}
\pdfoutput=1
\usepackage{multicol}
\usepackage{graphicx}
\usepackage{booktabs}
\usepackage{amssymb,bm,mathrsfs,bbm,amscd}
\usepackage[tbtags]{amsmath}
\usepackage{lastpage}
\usepackage{multirow}

\begin{document}

\fancyhead[c]{\small Submitted to 'Chinese Physics C'} \fancyfoot[C]{\small 010201-\thepage}

\footnotetext[0]{Received XXX}

\title{A new method of energy calibration of position-sensitive silicon detector\thanks{Supported by the ‘100 Person Project’ of the Chinese Academy of Sciences and the National Natural Science Foundation of China (Grant Nos.11405224 and 11435014)}}

\author{%
        SUN Ming-Dao(孙明道)$^{1,2,3}$%
\quad HUANG Tian-Heng(黄天衡)$^{1}$%
\quad LIU Zhong(刘忠)$^{1;1)}$\email{liuzhong@impcas.ac.cn}\\
\quad DING Bing(丁兵)$^{1}$%
\quad YANG Hua-Bin(杨华彬)$^{1,2,3}$%
\quad ZHANG Zhi-Yuan(张志远)$^{1}$\\
\quad WANG Jian-Guo(王建国)$^{1}$%
\quad MA Long(马龙)$^{1}$%
\quad YU Lin(郁琳)$^{1,2}$%
\quad WANG Yong-Sheng(王永生)$^{1,2}$\\
\quad GAN Zai-Guo(甘再国)$^{1}$%
\quad ZHOU Xiao-Hong(周小红)$^{1}$%
}
\maketitle

\address{%
$^1$ Institute of Modern Physics, Chinese Academy of Sciences, Lanzhou 730000, China\\
$^2$ University of Chinese Academy of Sciences, Beijing 100049, China\\
$^3$ School of Nuclear Science and Technology, Lanzhou University, Lanzhou 730000, China
}

\begin{abstract}
An improved method of energy calibration of position-sensitive silicon detector is presented. Instead of a parabolic function used in the traditional method, a new function describing the relation of position and energy is introduced and better energy resolution is achieved. Using the new method, the energy resolution of the 8.088 MeV $\alpha$ decay of $^{213}$Rn is determined to be about 87 keV (FWHM), which is better than the result of the traditional method, 104 keV (FWHM). In addition, different functions can be tried in the new method, which makes the calibration of detectors with various performances possible.

\end{abstract}



\footnotetext[0]{\hspace*{-3mm}\raisebox{0.3ex}{$\scriptstyle\copyright$}2013
Chinese Physical Society and the Institute of High Energy Physics
of the Chinese Academy of Sciences and the Institute
of Modern Physics of the Chinese Academy of Sciences and IOP Publishing Ltd}%

\begin{multicols}{2}

\section{Introduction}

Fusion-evaporation reactions are generally used in the synthesis of heavy nuclei. Evaporation residues are separated from primary beam and products of transfer reaction by the electromagnetic separation device \cite {lab1}. Subsequently they are implanted into position-sensitive silicon detector. New isotopes are usually identified by establishing $\alpha$ decay chains leading to the known transitions. The position-sensitive silicon detectors (PSSD) are extensively applied in the experimental studies of heavy nuclei, where position and time correlation measurements between implanted nuclei and subsequent $\alpha$ decays are performed. The PSSD used in the present work is a type of X1-300 manufactured by Micron corporation (UK). This detector is ion-implanted with an active area of 50 mm $\times$ 50 mm and a thickness of 300 $\mu$m. The silicon detector is fully depleted when a reverse bias of about 30 V is applied. One surface of the detector, with a well-distributed resistive layer on it, is divided into 16 strips, 3.125 mm wide for each strip. Signals are output from the two ends of each strip. The other side of the detector evaporated with a thin layer of Aluminum is used to get the total signal of the silicon. In the present work, three PSSD detectors are installed side by side to form an implantation detector.

In the following two sections, a new method of energy calibration of PSSD will be introduced, and then the calibration results of the new method and traditional method \cite {lab2,lab3} will be compared through some long-lived $\alpha$ emitters produced in the $^{20}$Ne+$^{209}$Bi reaction.

\section{Energy and position calibration of PSSD}

Signals output from the two ends of each strip are amplified in preamplifier and main amplifier before input into ADC (Analog to Digital Converter). The position of an event along the strip is determined by the resistive charge division \cite {lab4}, while the resolution of position depends on the performance of detector and noise from electronics. The energy registered in one strip is the sum of energy output from the two ends of the strip. The position perpendicular to the strip is given by the strip number.

In Fig. 1, two-dimensional spectrum of signals from the two ends of a strip for external $\alpha$ sources is plotted. The sources are: $^{239}$Pu (5.157 MeV), $^{241}$Am (5.486 MeV) and $^{244}$Cm (5.805 MeV) \cite {lab5}, which correspond to three groups of events in the middle of Fig. 1.

Three abnormal phenomena are observed in Fig. 1. At first, each of the three groups of events is not linear distributed but curve. This is mainly caused by the mismatch between input impedance of preamplifier and resistance of our detector. Events in the center of strip have much longer rise time than those in the two ends of strip \cite {lab6,lab7,lab8}. As a result, the ballistic deficit is much more serious for the events in the center of strip. Additionally, some other reasons may also lead to this distortion \cite {lab9}. This distortion can be reduced by an appropriate design of preamplifier. Secondly, the deviation caused by noise is position dependent. Three types of noise are presented for a similar detector in reference \cite {lab10}, which are the noise from the input transistor of amplifier, thermal noise from the resistive layer and shot noise produced by the leakage current of the detector which is mainly caused by radiation damage. The amplifier noise and shot noise are larger at the ends of the strip than at the center of it \cite {lab10}. At last, the data points are limited to a finite sector region in Fig. 1, but not extend to the coordinate axes. It results from the finite input impedance of preamplifier and can be restored by a simple relation \cite {lab11}. This situation can be improved by an appropriate design of preamplifier and the increase of integral time of main amplifier. However, the increasing of integral time may result in worsening accuracy of position measurement. A simple model is developed to simulate these phenomena. The simulated results can reproduce the experimental data curve.

Due to the first distortion mentioned above, a linear calibration of energy is not enough. In order to improve the result of linear calibration, parabolic fitting is proposed in the traditional method, in which the systematic variations of vertex and curvature of the parabola with $\alpha$ energy must be considered \cite {lab2,lab3}. A new method of the improvement, which could be applied more universally to silicon detectors based on the resistive charge division, is presented below.

\subsection{Linear calibration of energy}

The energies $E_{1}$ and $E_{2}$ extracted from the two ends of a strip are proportional to their ADC amplitudes $x_{1}$ and $x_{2}$, respectively,
\begin{equation}
\label{eq1}
E_{1} = a_{1}x_{1}+b_{1},
\end{equation}
\begin{equation}
\label{eq2}
E_{2} = a_{2}x_{2}+b_{2}.
\end{equation}
The total energy is
\begin{equation}
\label{eq3}
E = E_{1}+E_{2}.
\end{equation}
Then, a relation of $E$ and $x_{1}$, $x_{2}$ can be obtained
\begin{equation}
\label{eq4}
x_{2} = -\frac{a_{1}}{a_{2}}x_{1}+\frac{1}{a_{2}}E-\frac{b_{1}+b_{2}}{a_{2}}.
\end{equation}

\begin{center}
\includegraphics[width=8cm]{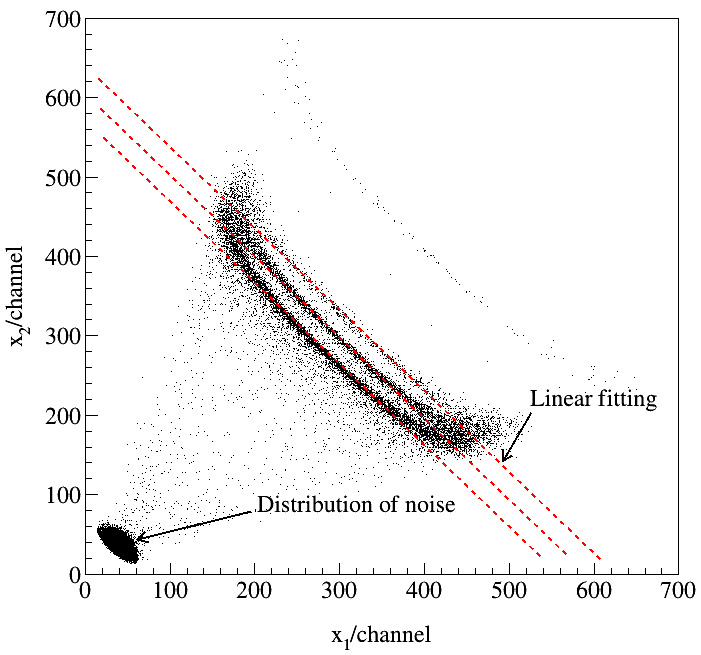}
\figcaption{\label{fig1}    A representative two dimensional spectrum of the raw energy signals (ADC amplitudes) from the two ends of one strip. }
\end{center}
Three parameters $k_{0}$=-$a_{1}$/$a_{2}$, 1/$a_{2}$ and  $b_{sum}$=($b_{1}$+$b_{2}$)/$a_{2}$ can be determined by linear fitting of data points (Fig. 1), where $b_{1}$ and $b_{2}$ are parameters to be determined later. Then, the energy obtained by linear calibration can be written as
\begin{equation}
\label{eq5}
E = a_{2}(-k_{0}x_{1}+x_{2}+b_{sum}).
\end{equation}

It is supposed that the center of noise distribution (shown in the lower corner of Fig. 1), namely the maximum of noise distribution density, is situated at the middle of a strip \cite {lab2}. This assumption is confirmed by a test in reference \cite {lab10}. So the output energies $E_{1}$ and $E_{2}$ corresponding to this center are equal. From Eqs (1) and (2), we have
\begin{equation}
\label{eq6}
a_{1}x_{n1}+b_{1} = a_{2}x_{n2}+b_{2},
\end{equation}
where $x_{n1}$ and $x_{n2}$ are the ADC amplitudes of noise at this center, then $b_{1}$ and $b_{2}$ can be obtained from $b_{sum}$ and Eq. (6).

\subsection{Position calibration}
As in reference \cite {lab3}, the position along a strip can be determined by the difference between energies $E_{1}$ and $E_{2}$ extracted from the two ends of the strip. The relative position $\eta$ is given as
\begin{equation}
\label{eq7}
\eta = \frac{E_{1}-E_{2}}{E_{1}+E_{2}},
\end{equation}
where $E_{1}$ and $E_{2}$ are the results of linear calibration of energy, shown in Eqs. (1) and (2). Relative position gives the position relative to the geometrical center of a strip, the range of which is between -1 and 1. Introduce Eqs. (1) and (2) into Eq. (7), we have
\begin{equation}
\label{eq8}
\eta = \frac{a_{1}x_{1}-a_{2}x_{2}+(b_{1}-b_{2})}{a_{1}x_{1}+a_{2}x_{2}+(b_{1}+b_{2})},
\end{equation}
The relation between real position and relative position is
\begin{equation}
\label{eq9}
P = \frac{L}{2}\eta,
\end{equation}
where $L$ is the length of strip, $L$ = 50 mm here.

\subsection{The improvement of linear calibration by new method}
The energy resolution can be improved by a new method based on the result of linear calibration.

The abnormal phenomena presented above can still be observed in Fig. 2, where the relation of relative position $\eta$ and energy E obtained from linear calibration is described. The events close to the ends of each strip (see Fig. 2) can’t be distinguished for different $\alpha$ sources, therefore only the events around the central part, encircled with colored lines (see Fig. 2), are chosen in the improvement of liner calibration. When an event occurs at one end of a strip, the output at the other end will be zero ideally, the range of $\eta$ will be 1 or -1. However, as mentioned above, due to the existence of input impedance of preamplifier, the range of relative position obtained by the resistive charge division is narrowed. It can be seen in Fig. 2 that the relative position $\eta$ spans approximately between -0.5 and 0.5.

\begin{center}
\includegraphics[width=8cm]{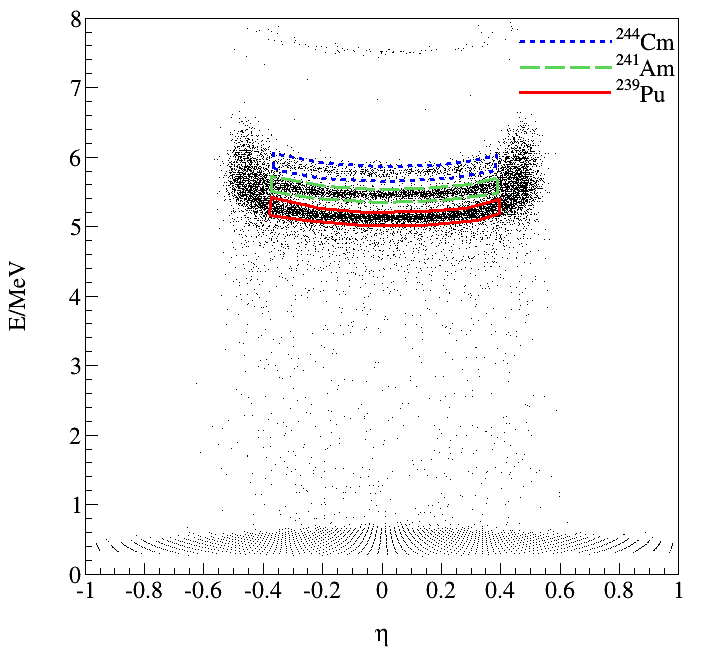}
\figcaption{\label{fig2}    Two dimensional spectrum of relative position $\eta$ and E, the energy obtained from linear calibration. Three groups of events circled by lines of different colors define good events of the three external $\alpha$ sources, which can be distinguished from each other.  }
\end{center}

\begin{center}
\includegraphics[width=8cm]{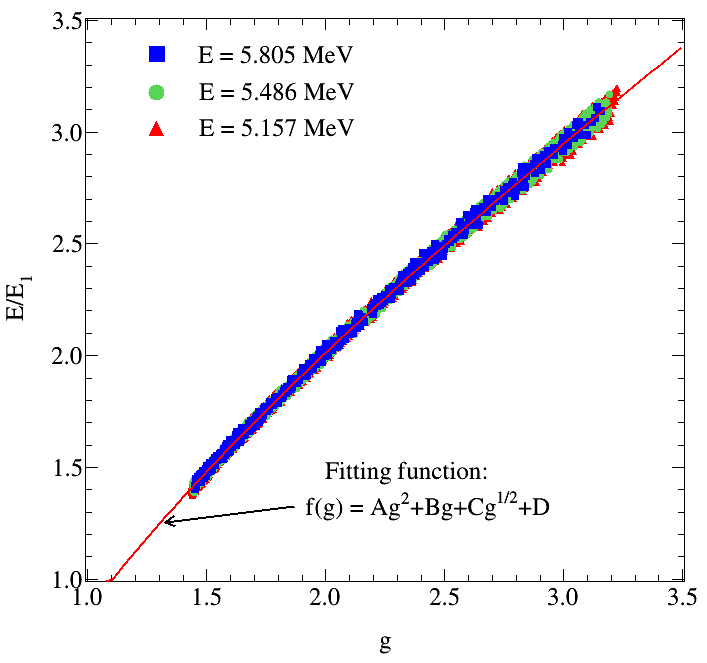}
\figcaption{\label{fig3}     Two dimensional spectrum of g and E/E1 for the selected events in Fig. 2, see text for explanation.  }
\end{center}

From Eqs. (3) and (7), we have
\begin{equation}
\label{eq10}
\frac{E}{E_{1}} = \frac{2}{1+\eta},
\end{equation}
make
\begin{equation}
\label{eq11}
g = \frac{2}{1+\eta},
\end{equation}
then
\begin{equation}
\label{eq12}
\frac{E}{E_{1}} = g.
\end{equation}
Eq. (12) indicates a linear relation between $E/E_{1}$ and $g$. But the data deviate from this simple relation.

\begin{center}
\includegraphics[width=8cm]{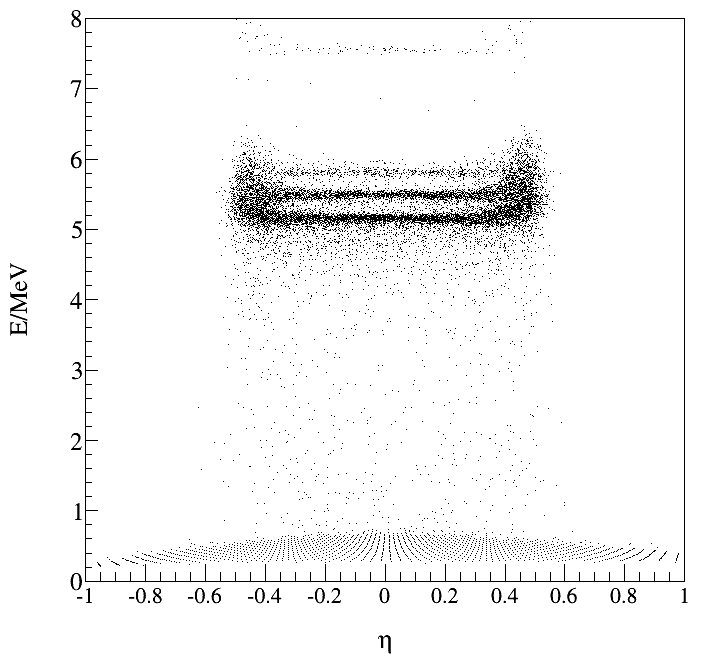}
\figcaption{\label{fig4}     Two dimensional spectrum of relative position $\eta$ and E, the energy improved by the new method.  }
\end{center}

In Fig. 3, the selected events of the three $\alpha$ sources in Fig. 2 are plotted in the $g-E/E_{1}$ coordinate system.
These three groups of events nearly overlap and follow a smooth curve. As a result, a new function, $f(g)$, is introduced
\begin{equation}
\label{eq13}
\frac{E}{E_{1}} = f(g)
\end{equation}
to describe the actual relation between $E/E_{1}$ and $g$.

Different functions have been tried fitting data points in Fig. 3, and a function with the following simple form
\begin{equation}
\label{eq14}
f(g) = Ag^{2}+Bg+Cg^{1/2}+D
\end{equation}
is found to be among the best. The results of linear calibration are improved by this function and new results are shown in Fig. 4 for all the data in Fig. 2.

There is some energy loss in the dead layer for the external $\alpha$ sources due to the existence of dead layer of the detector. So, for the internal $\alpha$ sources, the energy calibrated by the external $\alpha$ sources is larger than their real value. In order to reduce this difference, some long-lived residues in the $^{20}$Ne+$^{209}$Bi reaction are used to make a linear correction for the result obtained above. We have
\begin{equation}
\label{eq15}
E = [E_{1}(Ag^{2}+Bg+Cg^{1/2}+D)]K+M
\end{equation}
where $K$ and $M$ are the parameters of this linear correction.

\section{Comparison of the calibration results obtained by new method and traditional method}

This new method has been applied in the experimental study of the $^{20}$Ne+$^{209}$Bi reaction performed on the gas filled separator SHANS in Lanzhou \cite {lab12}. The beam energies were 111.4 MeV and 122.6 MeV for $^{20}$Ne.

\begin{center}
\includegraphics[width=8cm]{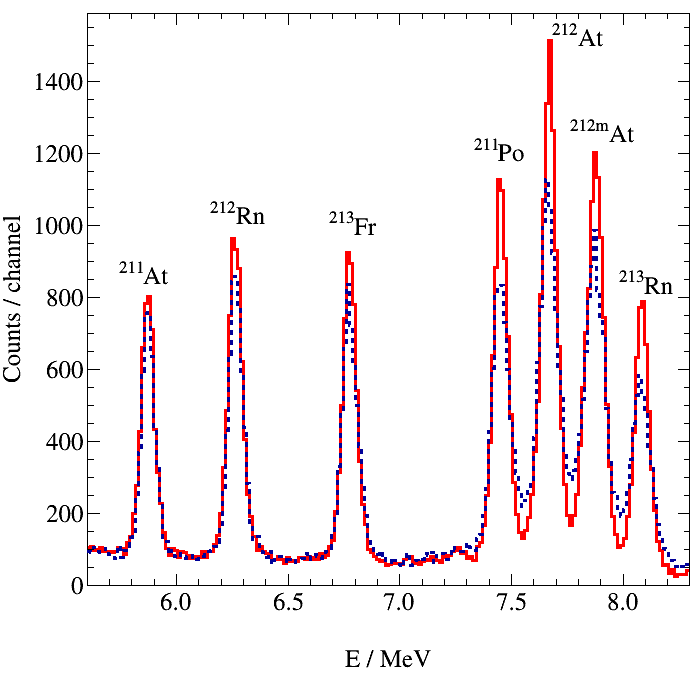}
\figcaption{\label{fig2}    A comparison between calibrated results by new method (red line) and traditional method (dashed blue line). }
\end{center}


\begin{center}
\tabcaption{ \label{tab2}  Comparison of energy resolutions}
\footnotesize
\begin{tabular*}{80mm}{@{\extracolsep{\fill}}cccc}

\toprule
\multicolumn{1}{c}{\multirow {2}{*}{nuclei}} &\multicolumn{1}{c}{\multirow {2}{*}{$\alpha$ decay energy/MeV}} &\multicolumn{2}{c}{FWHM/keV}\\
\cline{3-4}
\multicolumn{1}{c}{} &\multicolumn{1}{c}{} &\multicolumn{1}{c}{new} &traditional\\

\hline
$^{211}$At & 5.870 & 77 & 79\\
$^{212}$Rn & 6.264 & 78 & 80\\
$^{213}$Fr & 6.775 & 80 & 85\\
$^{211}$Po & 7.450 & 83 & 94\\
$^{212}$At & 7.679 & 88 & 98\\
$^{212m}$At & 7.837 & 99 & 114\\
$^{213}$Rn &	8.088 & 87 & 104\\
\bottomrule
\end{tabular*}%
\end{center}


The calibrated results by the new method and traditional method for the $\alpha$ decays of some isotopes produced in the $^{20}$Ne+$^{209}$Bi reaction are presented in Fig. 5, while the energy resolutions obtained are listed in Table 1. It is obvious that the new method produces better results than the traditional one, especially for the $\alpha$ decays with higher energies.

For the $\alpha$ decay (8.088 MeV) of the transfer reaction residue $^{213}$Rn in the $^{20}$Ne+$^{209}$Bi reaction, the energy resolution is determined to be 87 keV (FWHM) which is better than the result of traditional method, 104 keV. The position resolution is determined to be less than 2 mm (FWHM), similar to the result of traditional method. The position resolution is determined from the differences between the positions of short-lived implanted nuclei and those of the subsequent $\alpha$ decays. The resolution of position is not improved comparing with the result of traditional method, which may result from the big noise(seen in Fig. 1). Noise-reduction can improve the resolutions of energy and position.

\section{Summary}
A new method of energy calibration of position-sensitive silicon detector is introduced based on the observation that the distributions of data in the $g-E/E_{1}$ coordinate system are almost identical for different $\alpha$ sources (see Fig. 3). Compared with traditional method, the procedure is simpler and the obtained energy resolutions are better. Different functions can be tried in this new method making the calibration, therefore detectors with different performances can be calibrated by this new method, especially for detectors, the characteristic (namely the two dimensional spectrum of position and energy) of which violates parabolic shapes.

\end{multicols}

\vspace{-1mm}
\centerline{\rule{80mm}{0.1pt}}
\vspace{2mm}

\begin{multicols}{2}

\end{multicols}

\clearpage


\begin{thebibliography}{90}

\vspace{3mm}

\bibitem{lab1}Düllmann C E. Nuclear Instruments and Methods in Physics Research Section B, 2008, 266: 4123-4130


\bibitem{lab2}Folden C M. Development of odd Z projectile reactions for transactinide element synthesis, USA: University of California, 2004. 52-71

\bibitem{lab3}JIA G B, ZHANG Z Y et al. Nuclear Electronics \& Detection Technology, 2011, 31: 783-788

\bibitem{lab4}Alberi J L, Radeka V. IEEE Transactions on Nuclear Science, 1976, 23: 251-258

\bibitem{lab5}Browne E, Tuli J K. Nuclear data sheets for A = 235, Nuclear Data Sheets, 2014, 122: 205-293; Basunia M S.
   Nuclear data sheets for A =237, Nuclear Data Sheets, 2006, 107: 2323-2422; Singh B, Browne E. Nuclear data
   sheets for A = 240, Nuclear Data Sheets, 2008, 109: 2439-2499
\bibitem{lab6}Kalbitzer S, Melzer W. Nuclear Instruments and Methods, 1967, 56: 301-304
\bibitem{lab7}Doehring A, Kalbitzer S, Melzer W. Nuclear Instruments and Methods, 1968, 59: 40-44
\bibitem{lab8}Melzer W, Pühlhofer F. Nuclear Instruments and Methods, 1968, 60: 201-204
\bibitem{lab9}Kaufman S B, Wilkins B D, Fluss M J, Steinberg E P. Nuclear Instruments and Methods, 1970, 82: 117-121
\bibitem{lab10}Bassignana D et al. Nuclear Instruments and Methods in Physics Research Section A, 2013, 732: 186-189
\bibitem{lab11}LI Z Z, PENG H S et al. Nuclear Electronics \& Detection Technology, 1984, 4: 200-203


\bibitem{lab12}ZHANG Z Y. Experimental Study of the Superheavy Nuclide 271Ds on the Gas-Filled Recoil Separator in Lanzhou, China: Institute of Modern Physics, Chinese Academy of Sciences, 2012. 25-42


\end{thebibliography}
\end{document}